\documentclass[11pt]{article}

\usepackage[backend=biber]{biblatex}
\usepackage[utf8]{inputenc}
\usepackage[margin=1in]{geometry}
\usepackage{amsmath}
\usepackage{graphicx}
\usepackage{booktabs}
\usepackage{authblk} 
\usepackage{hyperref}
\usepackage{array}
\usepackage{indentfirst}
\usepackage{authblk}
\usepackage[skip=2pt]{caption}

\title{Bias in Surface Electromyography Features across a Demographically Diverse Cohort}

\author[1]{Aditi Agrawal*}
\author[1]{Celine John Philip*}
\author[2]{Giancarlo K. Sagastume}
\author[3]{Marcus A. Battraw}
\author[4,5,6]{Wilsaan M. Joiner}
\author[7]{Jonathon S. Schofield}
\author[4,8,9]{Lee M. Miller{$^{+}$}}
\author[6,9]{Richard S. Whittle{$^{+}$}}

\affil[1]{Graduate Group in Computer Science, University of California, Davis} 
\affil[2]{Department of Electrical and Computer Engineering, University of California, Davis} 
\affil[3]{Department of Mechanical and Mechatronic Engineering and Advanced Manufacturing, California State University, Chico,} 
\affil[4]{Department of Neurobiology, Physiology \& Behavior, University of California, Davis} 
\affil[5]{Department of Neurology, University of California, Davis} 
\affil[6]{Department of Physical Medicine and Rehabilitation, University of California, Davis} 
\affil[6]{Department of Mechanical \& Aerospace Engineering, University of California, Davis} 
\affil[8]{Department of Otolaryngology / Head \& Neck Surgery, University of California, Davis} 
\affil[9]{Center for Mind \& Brain, University of California, Davis} 
\date{}

\begin{document}

\maketitle

\noindent *$^{,+}$These authors contributed equally to this work.

\section*{Abstract}
Neuromotor decoding from upper-limb electromyography (sEMG) can enhance  human-machine interfaces and offer a more natural means of controlling prosthetic limbs, virtual reality, and household electronics. Unfortunately, current sEMG technology does not always perform consistently across users because individual differences such as age and body mass index, among many others, can substantially alter signal quality. This variability makes sEMG characteristics highly idiosyncratic, often necessitating laborious personalization and iterative tuning to achieve reliable performance. This variability has particular import for sEMG-based assistive devices and neural interfaces, where  demographic biases in sEMG features could undermine broad and fair deployment.

In this study, we explore how demographic differences affect the sEMG signals produced and their implications for machine learning-based gesture decoding. We analyze the data set provided by, in which we derive 147 common sEMG features extracted from 81 demographically diverse individuals performing discrete hand gestures. Using mixed-effects linear models and partial least squares (PLS) analysis, which take into consideration demographic variables (including age, sex, height, weight, skin properties, subcutaneous fat, and hair density), we identify that 33\% (49 of 147) of commonly used sEMG features show significant associations with demographic characteristics. These results may help guide the development of fair and unbiased sEMG-based neural interfaces across a diverse population.

\textbf{Keywords:} Surface electromyography, demographic sensitivity, feature extraction, mixed-effects models, neural interfaces, fairness in machine learning
\section{Introduction}
Decoding of hand gestures via upper limb surface electromyography (sEMG) enables individuals to control machines and computers by using their muscle activity as an input signal. The range of applications is broad, including prosthetic hands, virtual reality systems, and other consumer electronics \parencite{Atzori2014}. As surface EMG captures muscle electrical activity using sensors placed on the skin, it has the advantage of being non-invasive and relatively low cost.

However, one major challenge is that across individuals, these sEMG signals differ greatly, in part due to differences in anatomical and physiological characteristics such as body mass index (BMI), age-related variance in muscle activity, amount of subcutaneous tissue present beneath the skin surface, and many others. Despite the rapid evolution of wearable sEMG and machine learning (ML) technologies, the absence of publicly available and demographically diverse datasets and benchmarks that reflect the anatomical, physiological, and contextual diversity of real users raises several challenges \parencite{salterEmg2poseLargeDiverse2024}. Machine learning models trained on skewed datasets that do not represent true distributions of certain demographic groups result in incorrect classification of those underrepresented groups \parencite{gowdaDatabaseUpperLimb2025}. Similarly, face recognition systems have been reported to have significant performance disparities across demographic groups when trained on biased datasets \parencite{klarePushingFrontiersUnconstrained2015}. Demographic and related biases potentially also affect sEMG-based interfaces if not accounted for in model development.

A plethora of literature suggests why sEMG signals may differ on the basis of demographics, e.g., how muscle strength declines with age and muscle contractile function changes with obesity \parencite{goodpasterLossSkeletalMuscle2006,tallisEffectsObesitySkeletal2018}. Some of these physiological changes may be reflected in sEMG signal features, yet which specific features are affected and how much the effect is still not well understood. In the light of this, it is important to gain understanding of these relationships.

With sEMG-based interfaces becoming a reality for both clinical and commercial applications, it is not only ethically but practically necessary that these technologies work equally well for all demographic groups. Studying demographic-sensitive features gives us ways to develop preprocessing methods, feature selection, and modeling techniques that can be used to reduce the demographic bias of the models.

These factors motivate the central question of this study: How do demographic factors such as age, sex, BMI, and skin/tissue properties influence sEMG signal features? To address this gap, this project aims to comprehensively investigate the effects of age, BMI, and physiological properties including skin hydration, skin elasticity, subcutaneous fat thickness, and hair density on sEMG signals, providing the foundation required to develop adaptive and fair sEMG models for the general population. The sEMG dataset described by \parencite{gowdaDatabaseUpperLimb2025} includes a large population with a broad demographic. We then analyze 147 commonly used sEMG features extracted from 81 individuals after Multivariate Imputation by Chained Equations (MICE) for missing demographic variables. Mixed-effects linear models and partial least squares (PLS) were used to incorporate the demographic characteristics while accounting for mean differences among individuals, gestures, and sEMG channels. Knowledge of which features are most correlated with demographics will help with best practice for feature selection and demographically normalised modeling, as well as advancing our understanding of sEMG physiology. Ultimately, the development of fair sEMG-based control systems will open doors for new innovations in accessibility and human-machine interfaces, leading to the next generation of assistive technologies.
\section{Data Collection}
All practices of human subject research for this study \parencite{gowdaDatabaseUpperLimb2025} are in compliance with the Declaration of Helsinki and approved by the Institutional Review Board of the University of California Davis. All individuals provided written informed consent, including consent for publication of deidentified data. A total of 91 healthy adults were recruited, representing a broad demographic spectrum with a wide age range of 18 to 92 years (mean age 53.53 years, SD = 24.37 years), consisting of 37 males and 54 females. The study included individuals with varying BMI, and notably, 52 individuals were over 60 years old, with 23 categorized as having high BMI, indicative of overweight/obese classification in the original dataset ($BMI \geq 30$ kg/m²). Inclusion criteria included the participants being 18 years of age or older and able to understand, read, and speak English. \parencite{gowdaDatabaseUpperLimb2025}.

sEMG signals were recorded using Delsys Trigno double-differential mini electrodes, which wirelessly transmitted data to a base station. The acquisition platform utilized an NI USB-6210 16-channel multifunction I/O device with 16-bit resolution and a maximum sampling rate of 250 kS/s, capable of recording sEMG data at 2000 Hz or 2148 Hz utilizing hardware filtering. Twelve electrodes were placed circumferentially on the subject’s dominant (typically right) forearm: eight were positioned around the bulk muscle belly approximately two-thirds proximally from the wrist, and four were placed around the wrist joint to obtain full coverage of the involved muscle groups. Moreover, auxiliary physiological and biometric data were collected, including self-reported demographic variables (age, height, weight, sex) and professional device measurements of anatomical measurements: skin elasticity (Delfin Elastimeter), skin hydration (Delfin Moisture MeterSC), and hair density measured in hairs/cm² at two locations: forearm anterior and wrist anterior (Aram Huvis API 202). Subcutaneous fat (skinfold thickness) was measured in millimeters using MEDCA body fat calipers at four arm locations: forearm anterior, forearm posterior, wrist anterior, and wrist posterior. The individuals performed 10 various hand movements in six test sessions, totaling 360 movements to ensure robust data collection. Time synchronization between visual cues and sEMG data recording was maintained through ZeroMQ sockets and the Lab Streaming Layer (LSL) library and with minimal processing on the recorded data, which was separated into trials corresponding to gesture performance windows.
\section{Methodology}
\subsection{Data Representation}
All analyses were conducted on the dataset as released by \parencite{gowdaDatabaseUpperLimb2025}. For each individual, data were stored in serialized Python Pickle (.pkl) files. Each file contained a dictionary with four entries: sEMG, Labels, Frequency, and Physiology.
The sEMG entry consisted of a three-dimensional NumPy array of shape (360, 12, T), corresponding to 360 gesture trials, 12 electrode channels, and T time samples per trial (T = 4296 for recordings sampled at 2148 Hz). The Labels array identified the gesture performed in each trial. The Frequency entry specified the sampling rate used during acquisition. The Physiology entry contained a nested dictionary of individual-specific demographic and physiological measurements, including age, height, weight, skin hydration, skin elasticity, subcutaneous fat thickness at four arm locations, and hair density at two locations. These variables were selected based on known effects on sEMG signal generation, propagation, and electrode–skin coupling.

Although the original dataset comprised 91 individuals, statistical analyses were conducted on data from 81 individuals after excluding 10 cases that were missing all four of the following physiological variables: height, weight, skin elasticity, and skin hydration. For the remaining participants, missing values were limited to one or two physiological variables. In these cases, missing data were imputed using a Multivariate Imputation by Chained Equations (MICE) procedure, leveraging the remaining available demographic and physiological measurements. Specifically, imputation was applied for 9 cases missing subcutaneous fat at one location, 4 cases missing hair density measurements at both locations, 1 case missing subcutaneous fat measurements at all four locations, and 1 case missing both height and weight.

\subsection{sEMG Feature Extraction}
From each sEMG trial, we extracted a set of 147 features that are commonly used for gesture decoding \parencite{abbaspourEvaluationSurfaceEMGbased2020}. We performed feature extraction separately for each electrode channel in each trial, and then combined the features from each channel to create trial-level feature vectors. We chose a wide range of features to capture various aspects of the sEMG signal structure that are important for gesture decoding and physiological differences.

To reduce the influence of gesture onset and offset transients, sEMG features were not computed over the full trial duration of 2 seconds. Instead, a centered temporal window was extracted from each trial, corresponding to samples 600–3400 of the 4000-sample recordings (i.e., the middle 70\% of each trial, approximately 1.3 s). This window captures the steady-state portion of the gesture while excluding initial and final transient artifacts. All reported sEMG features were computed from this middle window for each electrode channel. A comprehensive list of all 147 extracted features, including their abbreviations, mathematical definitions, and categorization, is provided in the Appendix \textbf{(see Tables A1-A4)}

\subsubsection{Time-Domain Features}
Time-domain features measured signal amplitude and temporal complexity. These included mean absolute value (MAV), root mean square (RMS), integrated absolute value (IAV), standard deviation, and variance. These metrics describe the overall intensity of muscle activation. Signal complexity was further evaluated by using waveform length (WL), zero crossings (ZC), slope sign changes (SSC), and peak-based measures. In addition, we computed higher-order statistics such as skewness, kurtosis, and entropy-based descriptors \parencite{abbaspourEvaluationSurfaceEMGbased2020,battrawSurfaceElectromyographyEvaluation2024,Chowdhury2013a}.

\subsubsection{Frequency-Domain Features}
Frequency domain features were extracted from the power spectrum of the signal of each trial. These included mean frequency (MNF), median frequency (MDF), and peak frequency (PKF). We also calculated spectral energy within continuous frequency bands, such as 10 Hz bins. These features describe how sEMG signal power spreads across different frequencies.

\subsubsection{Time–Frequency Features}
To capture the nonstationary characteristics of sEMG signals, we extracted features based on wavelet packet transform (WPT). For each wavelet packet node, we computed features like logarithmic RMS energy, relative energy, and normalized log energy measures. This allows for localized time-frequency analysis of muscle activation patterns.

\subsubsection{Inter-Channel Features}
Inter-channel coordination was quantified using Pearson correlation coefficients calculated between electrode channels. This method captures spatial relationships and distributed muscle activation across the forearm and wrist.

\subsection{Statistical Analysis}
All statistical analyses were performed in R (version 4.3.2; R Core Team, 2025).

\subsubsection{Mixed-Effects Linear Models}
Associations between demographics and each sEMG feature were investigated by means of mixed-effects linear models. Demographic and physiological variables were considered as fixed effects. To model cases where the same persons, gesture types and electrode channels were observed several times, a random intercept was added. The modeling framework thus enables one to distinguish the share of population-level demographic effects from those of subject-specific, gesture-specific, and channel-specific variability.

\subsubsection{Significance Testing and Multiple Comparisons}
For each feature–demographic pair, hypothesis testing was performed to evaluate whether the estimated association differed significantly from zero (two-tailed). Corresponding p-values quantify evidence against the null hypothesis of no association. Given the large number of statistical tests conducted across features and demographic variables, false discovery rate (FDR) correction was applied to control for multiple comparisons.

\subsubsection{Effect Size Estimation}
In addition to statistical significance, $\eta^2$ was computed as a measure of effect size. $\eta^2$ represents the proportion of variance in a given sEMG feature attributable to a specific demographic variable and provides a scale-independent measure of practical significance.

\subsubsection{Partial Least Squares Analysis}
Partial least squares (PLS) regression was used to characterize the multivariate covariance between the sEMG features and demographic/physiological variables. This complements the univariate mixed-effects results. Specifically, we fit a sparse model (sPLS), extracting latent components that maximize the covariance between linear combinations of features and demographics. We focus only on the first sPLS component as the only one with significant predictive performance (Q²); subsequent components did not yield significant Q² and were not interpreted.

Feature and demographic loadings from the leading component(s) were then visualized with a clustered image map (CIM), which groups features and demographic variables with similar loading profiles. This analysis identified co-varying groups of features that share common multivariate associations with demographic factors.

\subsubsection{Feature-Level Demographic Sensitivity Aggregation}
To synthesize demographic effects across the full feature set, model outputs were summarized into feature–demographic matrices of FDR-corrected p-values and $\eta^2$ effect sizes. For each sEMG feature, the number of significant demographic associations, cumulative and mean $\eta^2$ values, maximum observed $\eta^2$, and the most influential demographic predictors were computed. These aggregated metrics form the basis for ranking sEMG features according to their demographic sensitivity. Associations were considered practically significant only if they met both criteria of an FDR-corrected two-tailed p-value $<$ 0.05 and a partial $\eta^2$ $\geq$ 0.06.

\section{Results}

We present three analyses based on the mixed-effects model: (1) ranking sEMG features by their number of significant demographic associations, (2) comparing effect size distributions across demographic variables, (3) visualizing specific feature-demographic interaction patterns. We furthermore investigated how patterns of demographics covary with patterns of sEMG features by (4) examining multivariate associations using a Partial Least Squares (PLS) clustered image map.
 
109 feature-demographic pairs showed practical significance (FDR-corrected p $<$ 0.05, partial $\eta^2 \ge 0.06$). In Figure~1, we rank features by the number of significant associations to reveal those that were most broadly influenced by demographics.

\subsection{Features most sensitive to demographics}

\begin{figure}[htbp]
    \centering
    \includegraphics[width=0.9\textwidth]{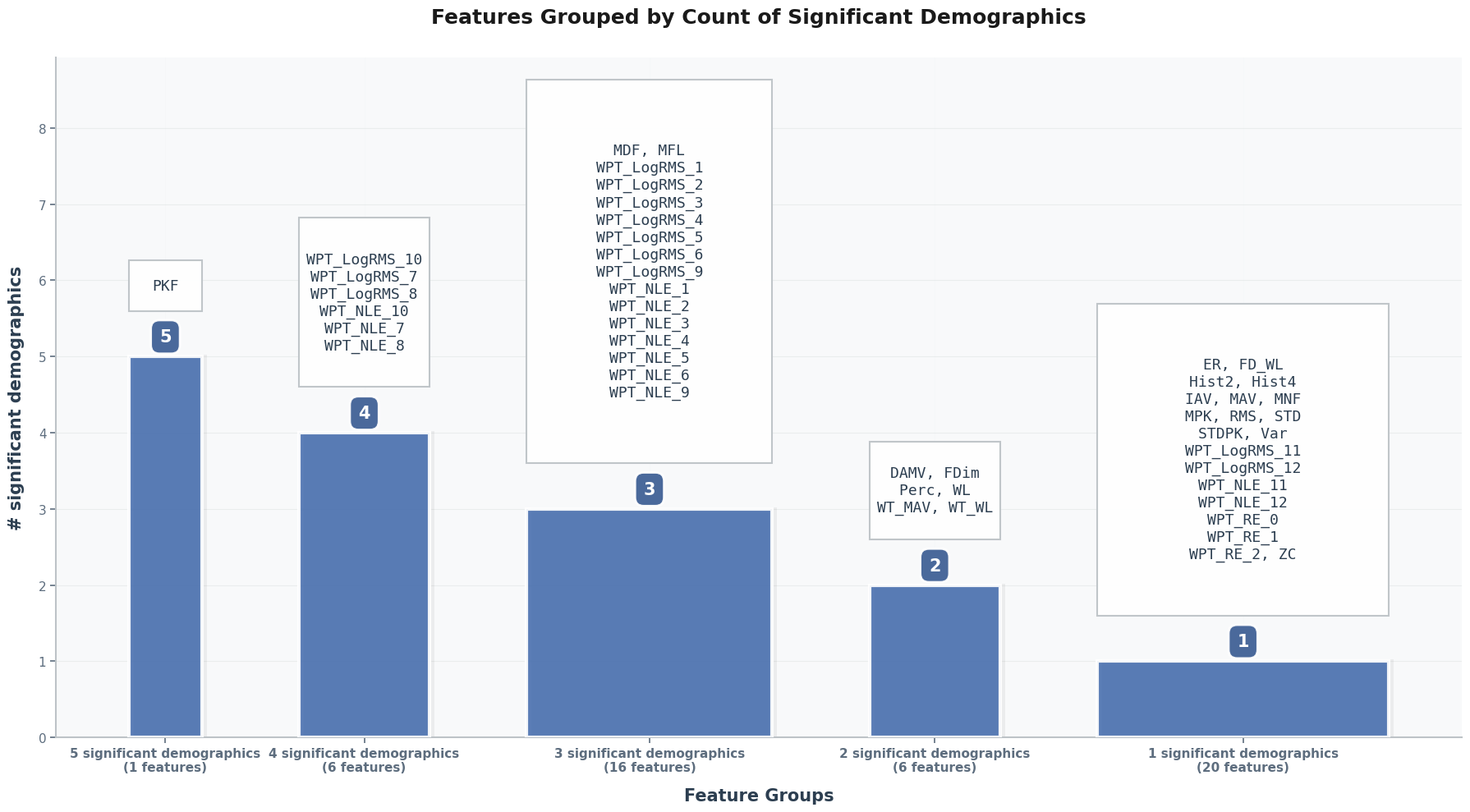}
    \caption{Top sEMG Features by number of significant Demographic associations}
    \label{fig:feature_rank}
\end{figure}

\vspace{-1.5em}
\begin{quote}
\small\itshape
sEMG features are grouped by the number of demographic variables with significant associations (FDR-corrected $p < 0.05$, partial $\eta^2 \ge 0.06$). Features are grouped into bars representing counts of 5, 4, 3, 2, 1, and 0 significant demographics, with feature names listed for each group. The width of each bar scales with the number of associated features. The only feature with 5 significant demographic associations is Peak Frequency (PKF). Six features (WPT\_LogRMS\_7, WPT\_NLE\_8, WPT\_NLE\_7, WPT\_LogRMS\_10, WPT\_LogRMS\_8, WPT\_NLE\_10) have 4 significant associations.
\end{quote}

In this section, we ranked the features by the number of demographics that had significant associations with them (Figure~1, bar plot of ``counts per feature''). Peak Frequency (PKF) stands out as the most demographically broadly-sensitive feature, with five significant demographic predictors (Age, Height, \texttt{Skin\_Hydration}, \texttt{Subcutaneous\_Fat\_1}, Sex). PKF represents the frequency at which the sEMG power spectrum reaches its maximum and shows associations with multiple demographic factors simultaneously.  

After PKF, the WPT features take a large part of the features ranking. These features come from a 4-level Symlet-5 wavelet packet decomposition, which divides the sEMG signal into 16 frequency bands; for each node, we calculate three metrics: \texttt{WPT\_LogRMS} (log root, mean, square of coefficients), \texttt{WPT\_RE} (relative energy), and \texttt{WPT\_NLE} (normalized log energy). Several \texttt{WPT\_NLE} and \texttt{WPT\_LogRMS} nodes (especially nodes 6--10) each reveal 3--4 significant demographic predictors with large effect sizes concentrated for Sex and \texttt{Subcutaneous\_Fat\_1/4}. 

These results highlight the sEMG features whose sensitivity to demographic changes is broadest, in the sense of having a higher number of significant demographic associations. Figure~2 additionally provides a summary of the effect size distributions for each demographic variable, reflecting their relative influence or strength.

\subsection{Distribution of effect sizes across demographics}

\begin{figure}[htbp]
    \centering
    \includegraphics[width=0.9\textwidth]{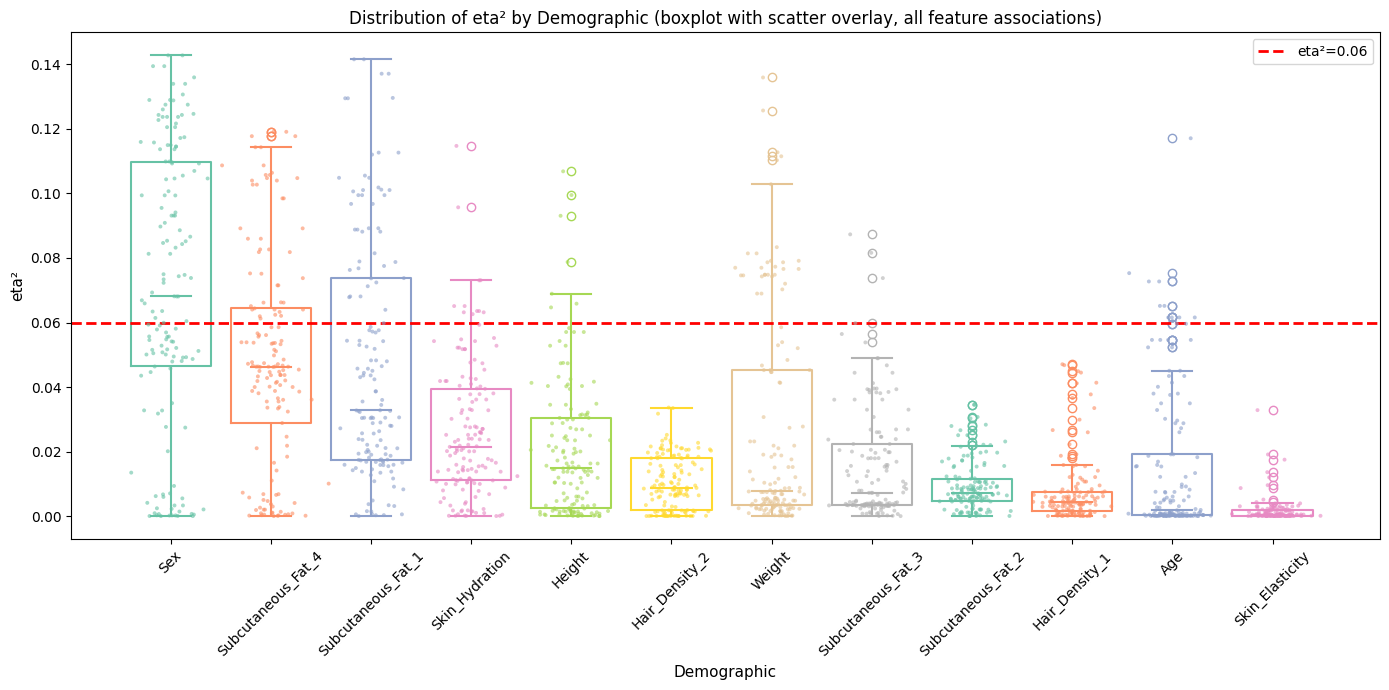}
    \caption{Distribution of Effect Sizes Across Demographics}
    \label{fig:distribution_effect_sizes}
\end{figure}

\vspace{-1.5em}
\begin{quote}
\small\itshape
Distribution of $\eta^2$ effect sizes across demographics for all 1,764 feature–demographic pairs, shown as boxplots with jittered scatter. Sex exhibits the largest median $\eta^2$ and the widest range, followed by Subcutaneous\_Fat\_4 and Subcutaneous\_Fat\_1, while other demographic variables show lower medians with only a few moderate-to-large effects.
\end{quote}

We summarized the effect sizes of 1,764 feature-demographic associations in Figure~2. The demographics in the plot are ranked according to their median effect size, thus revealing a distinct hierarchy. Sex is the demographic with the largest median $\eta^2$ (0.07) as well as the widest range. Several individual associations can be seen to exceed the $\eta^2 = 0.06$ threshold, and several outliers above 0.12 are also evident. The variables \texttt{Subcutaneous\_Fat\_4} and \texttt{Subcutaneous\_Fat\_1} come next with medians of about 0.045 and 0.03, respectively. In both cases, there are a lot of high ($\eta^2$) outliers, in particular, the largest effect in the dataset ($\eta^2 > 0.14$) for WPT node 8 with \texttt{Subcutaneous\_Fat\_1}.  

Although the medians for skin hydration and height are lower (0.02 and 0.015), they still show isolated associations with moderate to large effect sizes up to 0.10, which indicates the selective, feature-specific sensitivity. The rest of the demographics (Weight, \texttt{Subcutaneous\_Fat\_2/3}, Age, \texttt{Hair\_Density\_1/2}, \texttt{Skin\_Elasticity}) have medians close to zero and narrow interquartile ranges, with a few outliers only. In general, Figure~2 shows demographic effects to be very heterogeneous and feature-specific: a small number of demographics (sex and subcutaneous fat) are able to exert strong effects almost consistently across a large number of features, while the others affect only a small subset of features.

\subsection{Demographic patterns of effect size}

To complement the ranked demographic associations in Figure~2, Figure~3 reveals the particular feature-demographic interaction patterns that lead to these effects.

\begin{figure}[htbp]
    \centering
    \includegraphics[width=\textwidth]{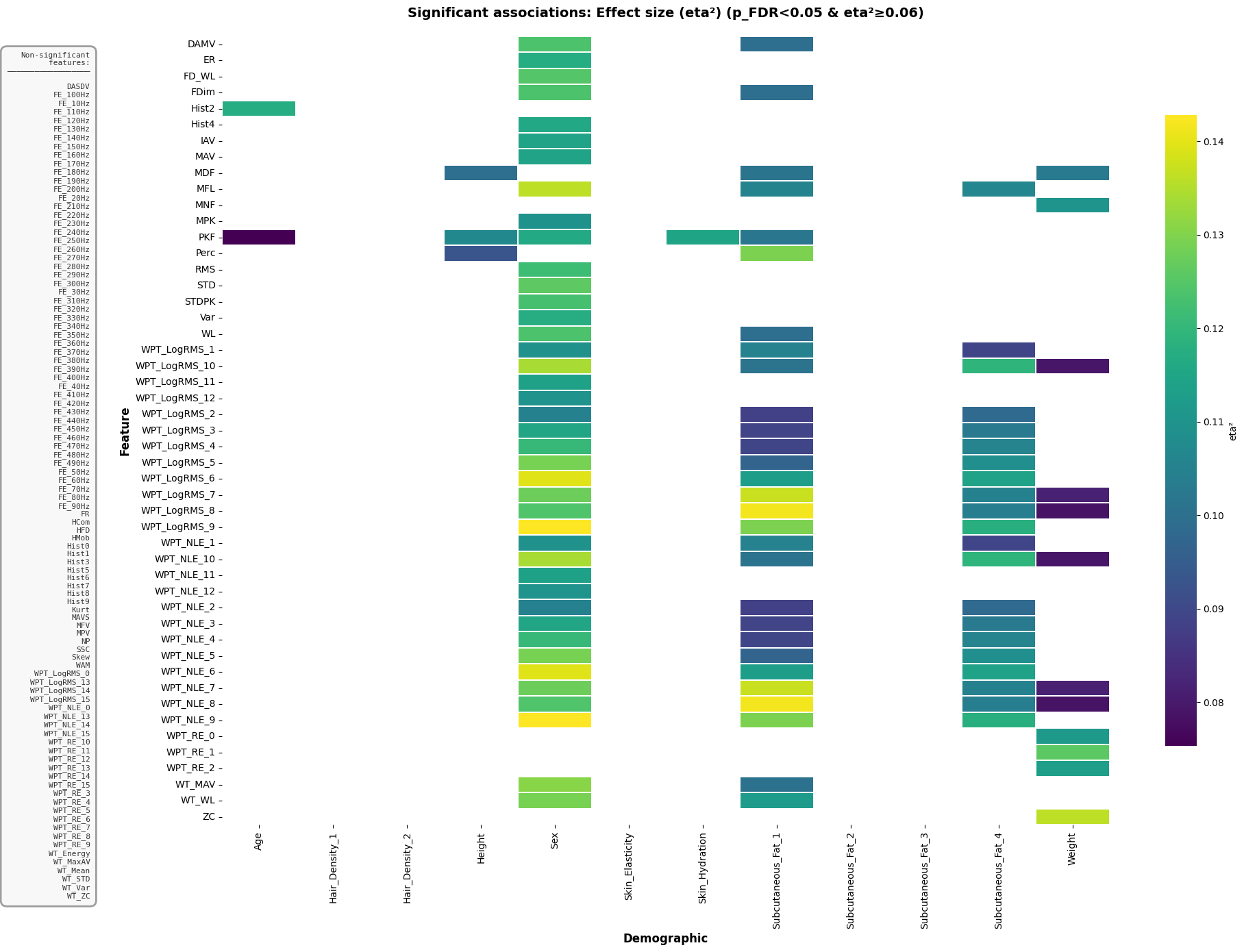}
    \caption{Significant Feature–Demographic Associations}
    \label{fig:Demographic_patterns_effectsize}
\end{figure}

\begin{quote}
\small\itshape
Heatmap of significant sEMG feature–demographic associations, displaying $\eta^2$ values for pairs passing FDR-corrected $p < 0.05$ and partial $\eta^2 \ge 0.06$. Clusters of large effect sizes are primarily observed for Sex and Subcutaneous\_Fat\_1/4 within frequency-domain and WPT features, whereas many time-domain and band-power features exhibit few or no significant associations.
\end{quote}

The $\eta^2$ heatmap of significant associations (Figure~3) demonstrates that the two main demographic determinants of variation in sEMG features are sex and subcutaneous fat (\texttt{Subcutaneous\_Fat\_1} and \texttt{Subcutaneous\_Fat\_4}). These demographics contribute the largest number of significant cells and the highest effect sizes ($\eta^2 \approx 0.08-0.14$). Thus, they are responsible for the bright clusters observed along the feature axis. On the other hand, Age, \texttt{Hair\_Density\_1/2}, Height, \texttt{Skin\_Elasticity}, \texttt{Skin\_Hydration}, \texttt{Subcutaneous\_Fat\_2}, and \texttt{Subcutaneous\_Fat\_3} only partially or not at all pass the dual significance thresholds of individual features, which implies that they have less broad influence at that level.

The major clusters in Figure~3 correspond to the wavelet packet transform (WPT) features, in particular, nodes 6--10. \texttt{WPT\_NLE}, \texttt{WPT\_RE}, and \texttt{WPT\_LogRMS} features originating from these nodes are most frequently found to be significantly associated with sex and subcutaneous fat measures, with a couple of contributions from weight. A number of frequency-domain features (e.g., kurtosis, MDF, MNF, NP, RMS, STDPK, WAM) likewise reveal significant associations with sex and subcutaneous fat. Time-domain and wavelet transform features (e.g., DAMV, \texttt{FD\_WL}, \texttt{WT\_MAV}, \texttt{WT\_STD}, \texttt{WT\_ZC}) show weaker but still moderate and consistent effects.

In contrast, frequency-band energy features (\texttt{FE\_10Hz--FE\_490Hz}) and FR (frequency ratio) fail to yield any cells that survive correction, indicating that these band power measures are relatively consistent and less influenced by demographic differences as per the criteria used.

Since features of the same WPT node represent different but closely related aspects of the same underlying frequency band, they have very similar demographic sensitivity profiles in Figures~1 and~3.

To determine the extent to which demographic variables are contributing compared to the other sources, the variance decomposition was also analyzed. The analysis of the variance explained in the mixed models showed a large difference between the conditional $R^2$ and the marginal $R^2$ measures. The conditional $R^2$ was large in magnitude (median = 0.65, range: 0.17--0.89) and indicated that the total variance was largely explained by the subject, gesture, and channel variables. The marginal $R^2$ was small in magnitude (median = 0.05, range: 0.002--0.17) but had a large significance, suggesting that the demographic variables also explained a significant amount of the total variance.

\subsection{Multivariate PLS patterns}

\begin{figure}[htbp]
    \centering
    \includegraphics[width=0.5577\textwidth]{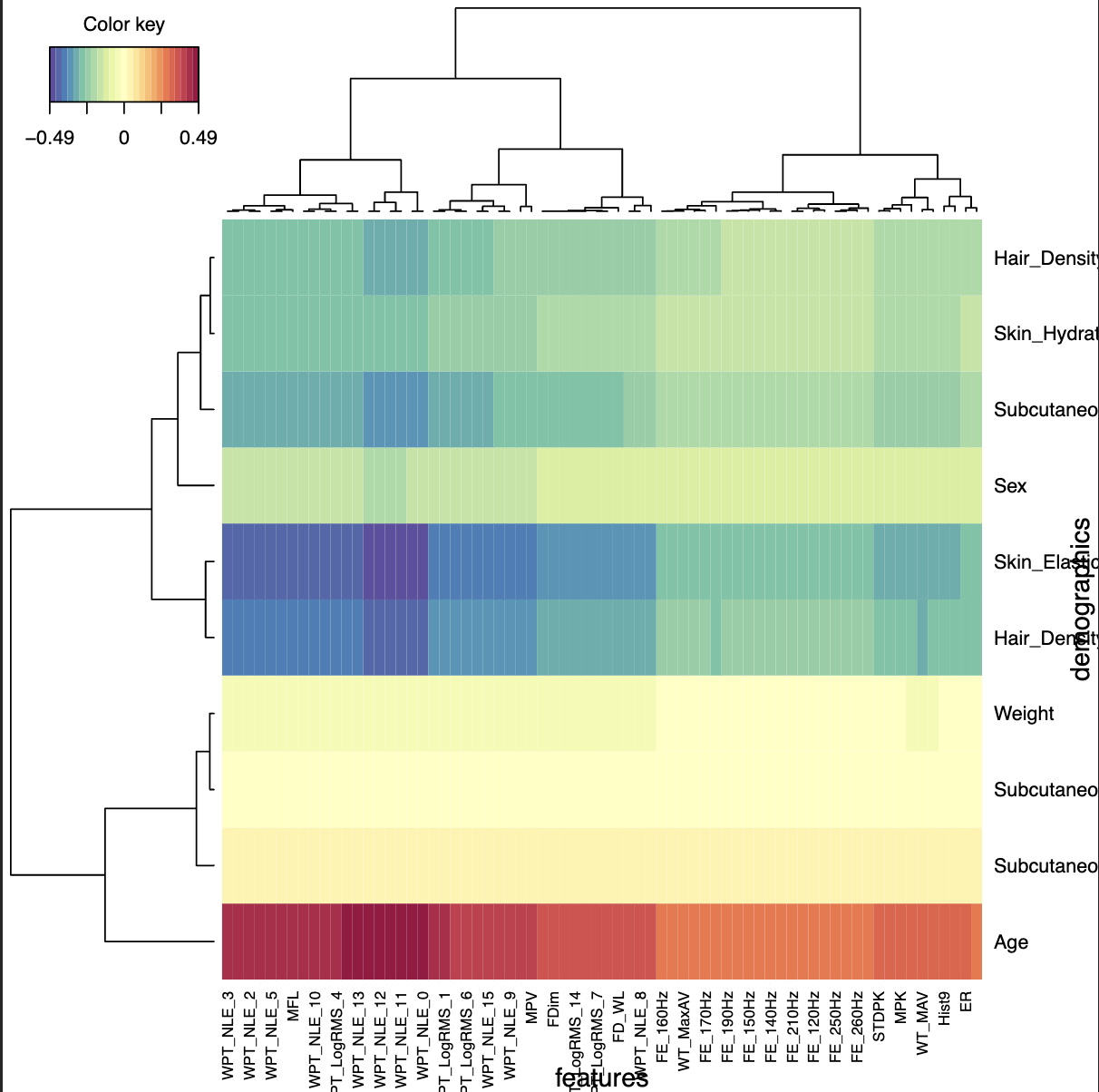}
    \caption{PLS clustered image map of sPLS component 1}
    \label{fig:PLS}
\end{figure}

\begin{quote}
\small\itshape
A Clustered image map displays the first sparse Partial Least Squares (sPLS) component features relating to sEMG on rows and demographic variables on columns. The color intensities represent the loadings or contributions of each feature–demographic pair on the component under examination after hierarchical clustering. We can see a large block of WPT and frequency-domain features seem to exhibit similar loading patterns with sex and subcutaneous fat. On the other hand, many time-domain features are scattered among different clusters, and hence, they show different loading expressions on the first component.
\end{quote}

To examine joint patterns of covariation between sEMG features and demographic variables, we analyzed the first sPLS component (the only component with significant $Q^2$). The CIM of this component is shown in Figure~4.

In addition to the mixed-effects models, we also applied a sparse Partial Least Squares (sPLS) analysis to the summed sEMG feature matrix and demographic matrix. This differs from the mixed effect model in that it captures patterns of demographics with patterns of sEMG features. The first sPLS component via a clustered image map (CIM) was plotted in Figure~4. The CIM places sEMG features as rows and demographic variables as columns, where the colors of the cells reveal how strongly each feature-demographic pair loads on the first sPLS component after hierarchical clustering is performed.

On the feature axis, several wavelet packet transform (WPT) features and multiple frequency-domain features appear together in a contiguous block. These features exhibit similar loading profiles on the first latent component, indicating that they covary as a group with specific demographic variables. In contrast to the loading pattern from that of the WPT and frequency-domain group, many basic time-domain features are located outside this main block, often in smaller clusters or more isolated positions, reflecting weaker multivariate associations.

On the demographic axis, the sex and subcutaneous fat measures (notably \texttt{Subcutaneous\_Fat\_1} and \texttt{Subcutaneous\_Fat\_4}) are clustered adjacent to the region of the CIM, containing the WPT and frequency-domain feature block, whereas other demographics, such as age, height, and skin properties, are arranged farther away or in smaller clusters. This arrangement shows that the sex and subcutaneous fat have a more similar pattern of loadings with WPT and frequency-domain features than with most time-domain features. Although age has a strong and relatively consistent loading across sEMG features on the first sPLS component, this global effect mostly limits the contribution of age to the clustering structure. On the other hand, sex and subcutaneous fat have specific loading patterns that result in the formation of the observed WPT and frequency-domain features clusters. Therefore, Figure~4, in general, gives a multivariate perspective that is still in line with Figures~1, 2 and~3 by identifying a group of WPT and frequency-domain features that are jointly patterned with sex and subcutaneous fat, while a large number of time-domain features are located in different parts of the clustering arrangement.

Together, the four visualizations (i) feature ranking by demographic associations (Figure~1), (ii) effect size distributions across demographics (Figure~2), (iii) feature demographic heatmap patterns (Figure~3), and (iv) the multivariate structure shown in the PLS clustered image map, illustrate that demographic factors influencing sEMG features are largely concentrated in the WPT and frequency-domain features, which is mainly attributed to sex and subcutaneous fat thickness with age contributing fewer but some of the strongest individual associations, whereas simple time-domain features are, for the most part, demographic-agnostic.
\section{Discussion}

The main finding of this research is that the demographic sensitivity of upper-limb sEMG features is not uniform: a relatively small set of WPT and frequency-domain features depends strongly on sex, subcutaneous fat thickness, and, in some cases, age, whereas basic time-domain features remain largely unchanged across demographic groups. This pattern directly informs which features are safest to use in demographic-agnostic models and which should be explicitly normalized or handled in a demographic-aware manner.

\subsection{Implications for Feature Selection}

This work has demonstrated that there are major differences in sensitivity to demographic and physiological variables in various sEMG feature types.

Specifically, the features related to nodes 6--10 from wavelet packet transforms, along with the peak frequency features, demonstrated robust associations with demographic variables, particularly sex and subcutaneous fat thickness. There were both greater numbers of demographic associations that were statistically significant and larger $\eta^2$ values than in other sets of features.

These results imply that using WPT and frequency-domain features in demographic-agnostic sEMG models should be done with caution. If WPT and frequency-domain features are used as inputs in models, demographic-normalized approaches or demographic-aware models may have to be considered to prevent biased model performance. In contrast, simple time-domain features such as mean absolute value (MAV), root mean square (RMS), standard deviation, and variance had only 1 significant association with the demographics. These are very robust and thus are attractive for consideration in scenarios where demographics are not known and cannot be accounted for in the modeling.

The observed consistency in demographic sensitivity profiles across features from the same WPT node suggests that these features can be combined or reduced without losing essential information. Node-level composites can be created by averaging or applying principal component analysis (PCA) over WPT\_LogRMS, WPT\_RE, and WPT\_NLE features from the same node, or by aggregating nodes 6--10 into a mid-high frequency composite. These combinations reduce redundancy and increase robustness, potentially yielding more interpretable frequency-band features that still capture the strong demographic sensitivity patterns observed in our results.

\subsection{Physiological Interpretation of Demographic Effects}

The demographic changes to the sEMG features that were observed have a biological basis and agree with the general principles of neuromuscular physiology and signal propagation. The effects of sex on the frequency-domain and time-frequency features were the most significant. It is well known that there are differences in muscle fiber composition between males and females \parencite{hicksSexDifferencesHuman2001}. In particular, the differences in the relative proportions of fast-twitch and slow-twitch fibers affect sEMG spectral characteristics \parencite{kupaEffectsMuscleFiber1995}.

These biological differences probably explain the large sex-related differences in PKF and WPT-based features. PKF's sensitivity to multiple demographic factors (Age, Height, Skin\_Hydration, Subcutaneous\_Fat\_1, Sex) likely reflects the combined influence of sex-related differences in muscle fiber composition, low-pass filtering by subcutaneous fat layers, changes in skin impedance with hydration, and body-size-related shifts in conduction velocity and muscle fiber length.

The thickness of the subcutaneous fat layer was very strongly correlated with frequency-domain and wavelet-based features. Fat tissue serves as a spatial low-pass filter, which attenuates the higher-frequency components of sEMG signals as they travel from the muscle to surface electrodes \parencite{Chowdhury2013a,kuikenEffectSubcutaneousFat2003}. Therefore, a thicker fat layer is expected to shift spectral energy to lower frequencies, a pattern that is in agreement with the demographic sensitivity of frequency-domain and WPT features to the fat thickness that was observed.

The interaction of sex and subcutaneous fat may thus account for the fact that these two variables were identified as the most significant demographic drivers. Muscle physiology affects the generation of the signal, while tissue composition affects the transmission of the signal, thus together determining the surface sEMG representation.

Some variables that may be considered to influence sEMG signals had unexpectedly small results in the analysis. Age had very limited associations despite the presence of known changes with age in muscle fiber composition, the properties of the motor units involved, and skin characteristics. Hair density and skin elasticity also had little influence despite the potential for the former and the latter variables to influence the quality of the electrode--skin interface. This indicates that the variables may influence the sEMG signal but to a far lesser extent compared to sex and body composition variables among the demographic characteristics of the study individuals.

The pattern of associations indicates that Subcutaneous\_Fat\_1 (forearm anterior) and Subcutaneous\_Fat\_4 (wrist posterior) have a greater effect compared to Subcutaneous\_Fat\_2 (forearm posterior) and Subcutaneous\_Fat\_3 (wrist anterior). Such a pattern could be related to the differences in the pattern of muscle activation during the execution of hand gestures or the anatomical differences in fat distribution. Nonetheless, these differences have yet to be explained.

\subsection{Implications for sEMG Model Development}

It is seen from the analysis of variance explained by these mixed-effects models that the values of conditional $R^2$ were very high and differed substantially from marginal $R^2$. High conditional $R^2$ shows that the majority of variance in sEMG features at the subject, gesture, and channel levels are explained by subject-specific, gesture-specific, and channel-specific effects, respectively. In contrast, modest marginal $R^2$ shows that demographic variables alone explain smaller though a statistically significant proportion of variance.

This pattern is important for sEMG model development. These results suggest that:  
i) Demographic variables supply useful contextual information, even if their explanatory power when considered alone is limited, and  
ii) Combination methods, which incorporate demographic normalization and subject-specific adaptation, can hence provide the most robust and fair performance in real sEMG systems.

\subsection{Multivariate PLS analysis}

The PLS clustered image map in Figure~4 gives a multivariate perspective that complements and is consistent with the mixed-effects models. It depicts that some WPT and frequency-domain features cluster together with sex and subcutaneous fat on the first sPLS component, whereas several time-domain features are located in different clusters with different loading patterns. The result is in line with the earlier finding that demographic sensitivity is mainly restricted to a small subset of WPT and frequency-domain features, whereas simple time-domain features, for the most part, remain unchanged across demographics.

\section{Limitations}

There are several limitations of this research that should be taken into consideration. First, the analysis is cross-sectional, which makes it impossible to identify causal relationships or track how demographic effects change over time. In other words, longitudinal data is necessary to be able to explain changes that occur due to aging or shifts in body composition that happen gradually. In addition, the extracted feature set contains feature redundancies. For example, many WPT-based features (e.g., LogRMS and normalized log energy for the same node) are highly correlated. Such redundancies can artificially increase the number of features that seem to be sensitive to demographic changes, though the problem is alleviated to some extent by the method of effect size aggregation. Also, even after correction for the false discovery rate (FDR), there is still a residual risk of false positives due to the large number of statistical tests performed. The use of a dual threshold combining statistical significance with a minimum effect size serves as a way of mitigating this risk, but it cannot completely rule it out.

Another limitation is that this analysis only considers demographic and physiological variables that have been measured in the dataset. Besides these, there are other factors of considerable influence, such as muscle mass, physical activity level, hand dominance, or even noise in electrode placement, which were not investigated here. We also did not do an analysis of gesture-specific or channel-specific demographic sensitivity that has not been carried out. Thus, gesture-specific or channel-specific demographic sensitivity remains an open question. While this work illuminates potential demographic bias in sEMG features, it does not address gesture decoding performance with those features. It may be that the most biased features tend to be more relevant for gesture decoding, or less so. Future work will determine which feature sets are both less biased and contribute to high decoding accuracy.

\section{Future Work}

The insights of this study open several avenues for future research. Longitudinal sEMG data would potentially allow the analysis of causation and dynamics, especially as it relates to the aging process and muscle physiology and composition changes. Future work could investigate whether there are differences in demographic sensitivity between particular gestures or electrode sites, allowing for more informed selection of features or placement of sensors. Based on the demographic-sensitive aspects that have been identified, demographic information could further be incorporated into the model through normalization of features using demographics, demographic-aware model architectures (for example, conditional or multi-branch networks), and fairness-constrained optimization goals.

Based on the observed robustness of basic time-domain features, we can explore the possibilities of designing new feature representations that preserve discriminative power while minimizing demographic sensitivity. More focused physiological research, for instance, simulation related to varying tissue thicknesses or activation rates, may help to validate the mechanisms causing the observed demographic effects. Demographic-sensitive sEMG features can be applied not only to gesture recognition but extended to muscle condition evaluation, rehabilitation, and geriatric neuromuscular evaluations as well.

\section{Conclusion}

This paper offers a detailed feature-level examination of demographic sensitivity in surface electromyography of upper limb gestures. This study, through mixed-effects modeling and effect-size analysis, has assessed 147 sEMG features across a demographically varied population and thus provides new insight into how population variability influences sEMG representations.

The findings indicate that sex and subcutaneous fat thickness are the two main demographic factors that drive the variation of sEMG features. The most significant changes for these factors have been found in frequency-domain and wavelet-based features. The multivariate PLS analysis confirms this pattern, showing that WPT and frequency-domain features also cluster together with these demographics on the first latent component. On the other hand, simple time-domain features seem to be less affected by demographic differences.

The results motivate the need for demographic-aware feature selection and model design to ensure fair and equitable sEMG-based neural interfaces. With sEMG technologies promising widespread clinical and commercial use, taking demographic variability into account is essential to ensure that the performance will be equal for all users.

The present analytical framework is a step towards the next research on demographic fairness in sEMG systems and may also be used for other biosignals and human-machine interface applications.

\section*{Acknowledgments}

This work was supported by Meta Platforms Technologies (Facebook Research) through the Ethical Neurotechnology program, with an award to L.M.M., and by the University of California, Davis School of Medicine Cultivating Team Science Award to L.M.M. We would like to thank Stephanie Naufel at Facebook Reality Labs for her valuable guidance and support. 
\section*{Code and Data Availability}

The sEMG dataset analyzed in this study is publicly available as part of the UCD-MyoVerse-Hand-1 dataset (\url{https://osf.io/3kzcb/overview}), released by Gowda et al.\ (2025). All feature extraction and statistical analysis code used in this work is available via a public GitHub repository and is sufficient to reproduce the reported results, figures, and tables:

\url{https://github.com/Aditii2112/Myoverse_EMG_analysis}

Processed feature matrices and summary statistics were generated directly from the released dataset using the described analysis pipeline.

\printbibliography

\newpage
\section*{Appendix}

\begin{table}[h!]
\caption{Table A1: Time-Domain (TD) Features}
\begin{tabular}{lp{10cm}}
\toprule
Abbreviation & Description \\
\midrule
MAV & Mean Absolute Value: Average of absolute signal amplitudes \\
STD & Standard Deviation: Signal variability measurement \\
Var & Variance: Measure of signal power spread \\
WL & Waveform Length: Cumulative absolute differences between samples \\
ZC & Zero Crossings: Count of signal sign changes \\
RMS & Root Mean Square: Quadratic mean representing signal power \\
NP & Number of Peaks: Count of local maxima in absolute signal \\
MPV & Mean Peak Value: Average amplitude of detected peaks \\
MFV & Mean Frequency Value: Average distance between consecutive peaks \\
SSC & Slope Sign Changes: Count of slope direction reversals \\
DAMV & Difference Absolute Mean Value: Mean of absolute differences \\
FDim & Fractal Dimension: Signal complexity measure \\
MFL & Mean Fractal Length: Logarithmic measure of signal fractal properties \\
HFD & Higuchi Fractal Dimension: Signal complexity estimation \\
Skew & Skewness: Measure of signal asymmetry \\
IAV & Integrated Absolute Value: Total signal energy \\
HMob & Hjorth Mobility: Frequency content indicator \\
HCom & Hjorth Complexity: Waveform complexity measure \\
ER & Energy Ratio: Total signal energy \\
DASDV & Difference Absolute STD Value: Variability in signal differences \\
WAM & Willison Amplitude: Count of large amplitude changes \\
MAVS & MAV Slope: Measure for multi-window analysis \\
Kurt & Kurtosis: Signal peakedness measurement \\
Perc & 75th Percentile: Value separating lowest 75\% of data \\
Hist0-9 & Histogram bins: Amplitude distribution across 10 bins \\
\bottomrule
\end{tabular}
\end{table}

\begin{table}[h!]
\caption{Table A2: Frequency-Domain (FD) Features}
\begin{tabular}{lp{10cm}}
\toprule
Abbreviation & Description \\
\midrule
FD\_WL & Frequency Domain Waveform Length \\
MNF & Mean Frequency: Power-weighted average frequency \\
MDF & Median Frequency: Midpoint of power spectrum \\
MPK & Mean Peak Amplitude: Average FFT peak magnitude \\
STDPK & Standard Deviation of Peak Amplitudes \\
FR & Frequency Ratio: Low vs high band power ratio \\
PKF & Peak Frequency: Dominant frequency component \\
FE\_10-490Hz & Frequency Energy: Sum of power in 49 discrete 10Hz bands \\
\bottomrule
\end{tabular}
\end{table}

\begin{table}[h!]
\caption{Table A3: Time-Frequency Domain (Wavelet) Features}
\begin{tabular}{lp{10cm}}
\toprule
Abbreviation & Description \\
\midrule
WT\_STD & Wavelet Coefficient Standard Deviation \\
WT\_Var & Wavelet Coefficient Variance \\
WT\_WL & Wavelet Waveform Length \\
WT\_Energy & Wavelet Energy Sum \\
WT\_MaxAV & Maximum Absolute Wavelet Coefficient \\
WT\_ZC & Wavelet Zero Crossings \\
WT\_Mean & Wavelet Coefficient Mean \\
WT\_MAV & Wavelet Mean Absolute Value \\
WPT\_LogRMS\_0-15 & Wavelet Packet Log-RMS for each of the 16 nodes \\
WPT\_RE\_0-15 & Wavelet Packet Relative Energy for each of the 16 nodes \\
WPT\_NLE\_0-15 & Wavelet Packet Normalized Log Energy for each of the 16 nodes \\
\bottomrule
\end{tabular}
\end{table}

\begin{table}[h!]
\caption{Table A4: Inter-Channel Features}
\begin{tabular}{lp{10cm}}
\toprule
Abbreviation & Description \\
\midrule
chX\_chY\_Cor & Pearson Correlation Coefficient: Measures coordination between all channel pairs \\
\bottomrule
\end{tabular}
\end{table}

\end{document}